\documentclass[sigconf]{acmart}
\usepackage{bm}
\usepackage{multirow}
\usepackage{algorithm}
\usepackage{algorithmic}
\def\model{DFEI}
\def\bmodel{\textbf{DFEI}}

\def\e{\bm{e}}
\def\x{\bm{x}} 
\def\X{\bm{X}} 

\def\Y{\bm{Y}} 
 
\def\v{\bm{v}}

\def\u{\bm{u}}
\def\s{\bm{s}}

\def\z{\bm{z}}

\def\L{\mathcal{L}}
\def\Domain{\mathcal{D}} 
\AtBeginDocument{%
  }

\acmConference[arxiv]{}{}{}
\acmISBN{978-1-4503-XXXX-X/18/06}

\begin{document}

\title{Large-Scale Multi-Domain Recommendation: an Automatic Domain Feature Extraction and Personalized Integration Framework}



\author{
Dongbo Xi$^{1,*}$,
Zhen Chen$^{1,*}$,
Yuexian Wang$^{1,2}$,
He Cui$^{1}$,\\
Chong Peng$^{1}$,
Fuzhen Zhuang$^{3,4}$, and
Peng Yan$^{1}$
}
\thanks{$*$ Corresponding authors: Dongbo Xi and Zhen Chen.}
\affiliation{
\institution{
$^1$Meituan\\
$^2$Shanghai Jiao Tong University, Shanghai 201109, China \\
$^3$Institute of Artificial Intelligence, Beihang University, Beijing 100191, China\\
$^4$SKLSDE, School of Computer Science, Beihang University, Beijing 100191, China\\
\{xidongbo,chenzhen06,cuihe02,pengchong,yanpeng04\}@meituan.com,\\
Wangyx1027@sjtu.edu.cn,
zhuangfuzhen@buaa.edu.cn
}
\country{}
}
\def\authors{Dongbo Xi, Zhen Chen, Yuexian Wang, He Cui, Chong Peng, Fuzhen Zhuang and Peng Yan}

\renewcommand{\shortauthors}{Xi and Chen, et al.}

\begin{abstract}
Feed recommendation is currently the mainstream mode for many real-world applications (e.g., TikTok, Dianping), in order to increase the overall daily active users of the application, it is usually necessary to model and predict user interests in multiple scenarios (domains) within and even outside the application. 
Multi-domain learning is a typical solution in this regard. 
The usual practice of multi-domain learning is to design shared and specific modules for each domain separately. 
While considerable efforts have been made in this regard, there are still two long-standing challenges:
(1) Accurately depicting the differences among domains using domain features is crucial for enhancing the performance of each domain. However, manually designing domain features and models for numerous domains can be a laborious task.
(2) Users typically have limited impressions in only a few domains. Extracting features automatically from other domains and leveraging them to improve the predictive capabilities of each domain has consistently posed a challenging problem.
In this paper, we propose an Automatic~ \textbf{D}omain~ \textbf{F}eature~ \textbf{E}xtraction~ and Personalized \textbf{I}ntegration~ (\bmodel)~ framework for the large-scale multi-domain recommendation. The framework automatically transforms the behavior of each individual user into an aggregation of all user behaviors within the domain, which serves as the domain features. Unlike offline feature engineering methods, the extracted domain features are higher-order representations and directly related to the target label.
Besides, by personalized integration of domain features from other domains for each user and the innovation in the training mode, the \model~framework can yield more accurate conversion identification. 
Experimental results on both public and industrial real-world Dianping feed recommendation datasets, consisting of over 20 domains, clearly demonstrate that the proposed framework achieves significantly better performance compared with state-of-the-art baselines.
Furthermore, we have released the source code of the proposed framework at 
https://github.com/xidongbo/DFEI.
\end{abstract}

\ccsdesc[500]{Computing methodologies~Multi-task learning}
\ccsdesc[500]{Information systems~Recommender systems}

\keywords{Multi-Domain Learning, Recommender System, Feed Recommendation, CTR Prediction}


\maketitle

\section{Introduction}
In feed recommendation, users can continuously browse items generated by an endless feed using their mobile phones.
Accurate feed recommendation is crucial for increasing the daily active users of real-world applications.
Especially with the increasing complexity of applications nowadays and the involvement of more and more feed recommendation scenarios (domains), it has always been a challenging problem to accurately depict the differences between domains and utilize the information of domains to improve the recommendation performance in each domain.

 The conventional approaches \cite{STAR, EDDA, HiNet} for multi-domain learning involve designing shared and specific modules for each domain separately. However, despite significant efforts in this area, there are still two long-standing challenges that need to be addressed.
Firstly, manually designing domain features and models for numerous domains can be a laborious task. This manual process not only consumes a significant amount of time and effort but also increases the risk of human error. Therefore, there is a need for an automatic approach that can extract domain features effectively and efficiently.
Secondly, users typically have limited impressions in only a few domains. 
When inferring in a new domain, the domain-related features of users are often missing.
This poses a challenging problem of extracting features automatically from other domains and leveraging them to improve the predictive capabilities of each domain. 
Existing methods struggle to overcome this challenge, and as a result, the performance of multi-domain recommendation systems remains sub-optimal.

To address these challenges, this paper proposes an Automatic Domain Feature Extraction and Personalized Integration (\model) framework for the large-scale multi-domain recommendation. The \model~ framework automatically transforms the behavior of each individual user into an aggregation of all user behaviors within the domain, which serves as the domain features. In contrast to traditional offline feature engineering methods, the extracted domain features are higher-order representations that are directly correlated with the target label. This approach not only minimizes the manual effort needed for feature engineering but also improves the performance of the feed recommendation.
Furthermore, the \model~ framework personally integrates domain features from other domains for each user. This integration, coupled with the innovation in the training mode, enables the framework to significantly enhance the recommendation performance.

To evaluate the effectiveness of the proposed framework, extensive experiments are conducted on both public and industrial real-world Dianping feed recommendation datasets, consisting of over 20 domains. The experimental results clearly demonstrate that the \model~ framework outperforms state-of-the-art baselines, achieving significantly better performance.

To summarize, the contributions of this paper are threefold:
\begin{itemize}
    \item The \model~ framework achieves automatic domain feature extraction and personalized integration for the large-scale multi-domain recommendation.
    \item The experimental results validate the effectiveness of the proposed framework, highlighting its superiority over existing approaches.
    \item The release of the source code further promotes collaboration and encourages future advancements in the field of multi-domain recommendation.
\end{itemize}

\section{Related Work}
This section offers a brief overview of some topics related to our study, including Multi-Task Learning (MTL) and Multi-Domain Learning (MDL). 

\subsection{Multi-Task Learning}
Multi-Task Learning (MTL) is a subfield of machine learning in which the model is trained simultaneously on multiple related tasks. The key idea is to leverage the commonalities and divergences across tasks to enhance the model's performance on each individual task, rather than training separate models for each task. Share-Bottom \cite{share_bottom} network is the most commonly used multi-task model which shares the bottom network to discern similarities and builds multiple task-specific networks at the top to identify differences. However, the hard parameter sharing approach may suffer from severe optimization conflicts when tasks are not highly correlated. Cross-Stitch \cite{Cross-Stitch} and Sluice Network \cite{Sluice} are proposed to learn unique combinations of the outputs for each task. MoE \cite{MoE} and MMoE \cite{MMoE} adopt gating networks to combine multiple experts shared at the bottom to capture different patterns in data, which makes it more effective when the tasks are loosely related. Based on MMoE, PLE \cite{PLE} separates expert networks into shared and task-specific to better tackle with seesaw phenomenon. Unlike preceding approaches, models such as ESMM \cite{ESMM}, ESM2 \cite{ESM2}, and AITM \cite{AITM} primarily concentrate on characterizing the relationship between output tasks. Concurrently, certain studies \cite{MetaBalance, PAPERec, AdaTask} have been undertaken to mitigate the issue of negative transfer from the vantage point of multi-task optimization.

\subsection{Multi-Domain Learning}
Multi-Domain Learning (MDL) has become an increasingly significant research area due to the prevalence of data collected from various domains in real-world applications \cite{confidence-weighted, whendodomainsmatter, HMoE}. MDL seeks to enable knowledge transfer across different domains to enhance learning outcomes in all domains simultaneously, distinguishing itself from Transfer Learning (TL) \cite{TLsurvey1, TLsurvey2}, which traditionally involves a unidirectional transfer of knowledge from a source to a target domain \cite{DDTCDR, CoNet}.

In recent years, Multi-Domain Learning has been applied to recommendation systems, with a focus on Multi-Domain Recommendation (MDR) \cite{HMoE, SAML, SAR-Net, AFT, STAR, PEPNet, HiNet, MARIA, MI-DPG, SAMD, M3REC, MuSeNet, M5, PLATE, HAMUR, EDDA, Meta-generator, HDN}. Inspired by Multi-task learning, HMoE \cite{HMoE} takes advantage of MMoE \cite{MMoE} structure to implicitly model distinctions and commonalities among different domains. SAML \cite{SAML} constructs domain-independent and domain-dependent features respectively through feature transformation and attention mechanism, and proposes a mutual unit to learn similarities between domains. SAR-Net \cite{SAR-Net} employs attention modules to transfer users' interests across domains and uses a mixture of experts for information extraction. AFT \cite{AFT} proposes a generative adversarial network framework to capture informative domain-specific features. STAR \cite{STAR} constructs a star topology with shared centered parameters and domain-specific parameters, and trains a single model to serve all domains. Moreover, some other works start to focus on performing multiple tasks in multiple domains simultaneously. For instance, PEPNet \cite{PEPNet} introduces both parameter personalization and embedding personalization to accurately capture user interests for multi-task recommendation in the multi-domain setting. AESM$^2$ \cite{AESM2} and HiNet \cite{HiNet} use a hierarchical MOE to model the commonalities and differences between different domains and tasks. HiNet pays special attention to how useful information from one domain can be for another domain.

However, all of the above models do not explicitly and automatically extract domain features for the large-scale multi-domain recommendation, which is crucial for enhancing the performance of each domain, and also do not explicitly consider the information utilization problem when users only have impressions in a few domains.

\begin{figure*}[!t]
\begin{center}
\includegraphics[width=1.0\linewidth]{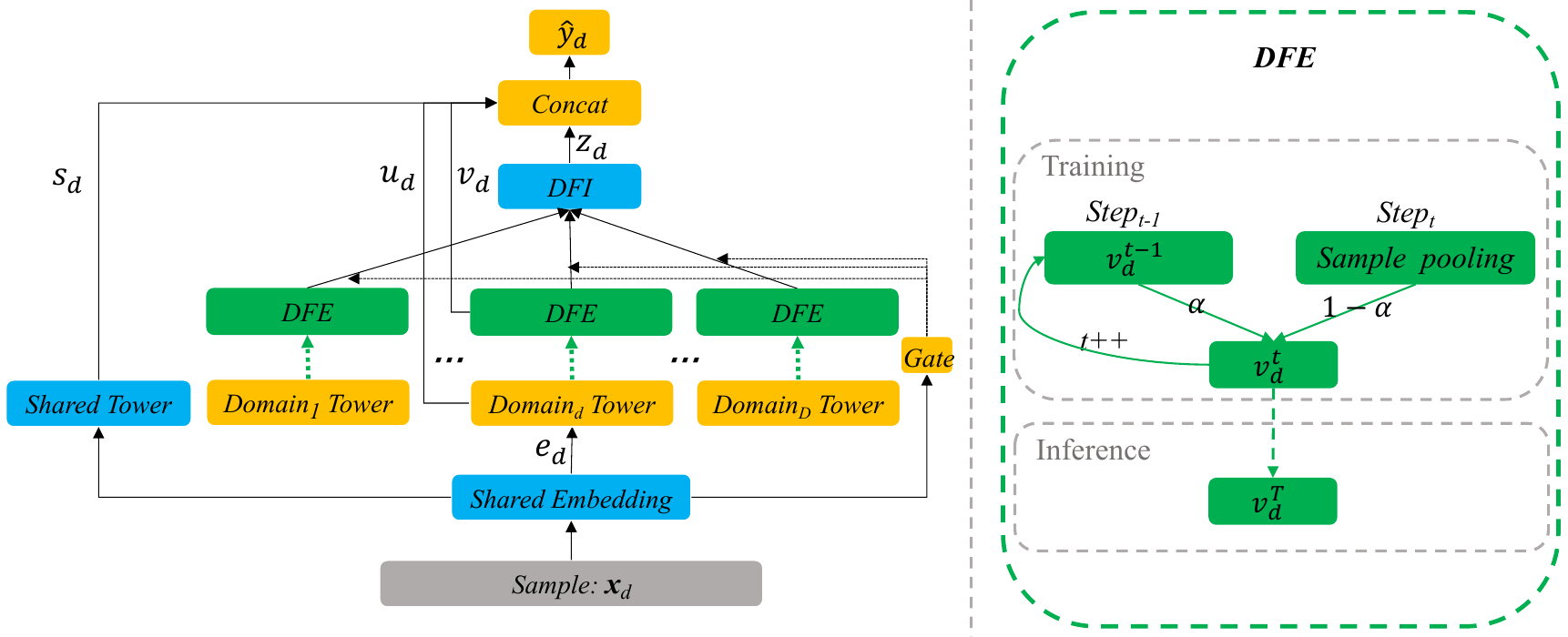}
\caption{The proposed Automatic Domain Feature Extraction and Personalized Integration (\model) framework.}
\label{fig:model}%
\end{center}
\end{figure*}%

\section{Methodology}
In this section, we first describe the problem formulation, and then we present the details of the proposed framework \model~ as shown in Figure \ref{fig:model}. 

\subsection{Problem Formulation}
As mentioned above, we focus on the large-scale multi-domain recommendation problem. Given the domain set $\Domain$, assuming the $d$-th domain sample is $\x_d$,
($\x_d\in\Domain_d, d = 1, 2, \cdots, D$). 
The multi-domain framework needs to predict the conversion probability $\hat{y}_d$ of each domain $d$ based on the input features $\x_d$:
\begin{equation}
\hat{y}_d = p(y_d=1|\x_d).
\end{equation}%

\subsection{Automatic Domain Feature Extraction and Personalized Integration}
As shown in Figure \ref{fig:model}, given the sample $\x_d$ in domain $\Domain_d$, we first encode all the features and then embed each of them to a low dimension dense vector representation, the output of the Shared Embedding layer is the concatenation of all embedding vectors:
\begin{equation}
\e_d = E(\x_d),
\label{eqt:e}
\end{equation}%
where $E$ denotes the embedding and concatenation operations. 
By sharing the same embedding vectors across all domains, several benefits can be achieved. Firstly, it helps to reduce the number of model parameters for the large-scale multi-domain recommendation. Secondly, it enhances information sharing between different domains, thereby improving overall performance.

Next, shared and specific modules are adopted to extract the shared and specific representation, respectively:
\begin{equation}
\s_d = f(\e_d ),~
\u_d = f_d(\e_d ),
\label{eqt:f}
\end{equation}%
where $f(\cdot)$ is the Shared Tower, and $f_d(\cdot)$ is the $\Domain_d$ Tower as shown in Figure \ref{fig:model}.
It should be mentioned that designing different shared and specific towers is not the focus of this paper as we aim at designing an Automatic Domain Feature Extraction and Personalized Integration modules to improve the overall performance for the large-scale multi-domain recommendation.

\textbf{DFE}:
Domain features can accurately depict the differences among domains, which are crucial for improving the performance of each domain.
Although the domain is static, users within the domain are dynamic, and the domain feature should be represented by the behavior of users within the domain. In the training phase, the Automatic Domain Feature Extraction computes the domain feature $\v_d^t$ at training step $t$ as:
\begin{equation}
\v_d^t = \alpha * \v_d^{t-1} + (1 - \alpha) * pooling(\u_d^t),
\label{eqt:avg}
\end{equation}%
where $\v_d^{t-1}$ is output of training step $t-1$, $pooling(\u_d^t)$ is the mean-pooling of the samples at training step $t$ during mini-bath training, $\v_d^0=pooling(\u_d^0)$, and $\alpha$ defines the decay coefficient.
Intuitively, the domain feature can be understood as the moving averages of all samples throughout the training process.
It transforms the behavior of each individual user into an aggregation of all user behaviors within the domain. 
In contrast to traditional offline feature engineering methods, the extracted domain features are higher-order representations that are directly associated with the target label during the training process.
In the inference phase, the domain feature will directly adopt the final trained state, i.e., $\v_d = \v_d^T$.
\begin{table*}[!t]
  \centering
  \caption{Basic information of the industrial dataset.}
  \resizebox{\textwidth}{!}{
    \begin{tabular}{c|ccccccccccccccccccccc}
    \toprule
    \textbf{Domains} & \textit{\textbf{\#A1}} & \textit{\textbf{\#A2}} & \textit{\textbf{\#A3}} & \textit{\textbf{\#A4}} & \textit{\textbf{\#A5}} & \textit{\textbf{\#A6}} & \textit{\textbf{\#A7}} & \textit{\textbf{\#A8}} & \textit{\textbf{\#A9}} & \textit{\textbf{\#A10}} & \textit{\textbf{\#A11}} & \textit{\textbf{\#A12}} & \textit{\textbf{\#A13}} & \textit{\textbf{\#A14}} & \textit{\textbf{\#A15}} & \textit{\textbf{\#A16}} & \textit{\textbf{\#A17}} & \textit{\textbf{\#A18}} & \textit{\textbf{\#A19}} & \textit{\textbf{\#A20}} & \textit{\textbf{\#A21}} \\
    \midrule
    \textbf{\#Train} & 2,514K & 1,881K & 83K   & 18K   & 83K   & 49K   & 328K  & 832K  & 213K  & 51K   & 146K  & 15K   & 292K  & 91K   & 9K    & 20K   & 66K   & 28,625K & 35K   & 2,135K & 127K \\
    \textbf{\#Validation} & 313K  & 235K  & 11K   & 2K    & 10K   & 6K    & 41K   & 104K  & 27K   & 6K    & 18K   & 2K    & 37K   & 11K   & 1K    & 2K    & 8K    & 3,587K & 4K    & 266K  & 16K \\
    \textbf{\#Test} & 314K  & 235K  & 11K   & 2K    & 10K   & 6K    & 41K   & 104K  & 27K   & 6K    & 18K   & 2K    & 37K   & 11K   & 1K    & 3K    & 8K    & 3,579K & 4K    & 267K  & 16K \\
    \midrule
    \textbf{CTR(\%)} & 1.09  & 1.9   & 2.13  & 3.09  & 3.47  & 3.59  & 3.67  & 3.83  & 4.6   & 5.39  & 5.4   & 5.76  & 5.98  & 6.46  & 6.79  & 6.98  & 7.46  & 7.58  & 7.84  & 7.86  & 9.45 \\
    \bottomrule
    \end{tabular}%
    }
  \label{tab:industry_dataset}%
\end{table*}%

\begin{table}[!t]
    \centering
    \caption{Basic information of the public dataset.}
    \resizebox{\linewidth}{!}{
    \begin{tabular}{c|cccccc}
    \toprule
    \textbf{Domains} & \textit{\textbf{\#B1}} & \textit{\textbf{\#B2}} & \textit{\textbf{\#B3}} & \textit{\textbf{\#B4}} & \textit{\textbf{\#B5}} & \textit{\textbf{\#B6}}\\
    \midrule
    \textbf{\#Train}  & 857K & 2,615K & 148K & 326K & 5K & 50K \\
    \textbf{\#Validation} & 209K & 671K & 35K & 77K & 2K & 17K \\
    \textbf{\#Test} & 1,341K & 4,431K & 219K & 493K & 9K & 117K\\
    \midrule
    \textbf{CTR(\%)} & 12.00 & 44.54 & 39.09 & 56.05 & 21.34 & 18.10\\
    \bottomrule
    \end{tabular}
    }
    \label{tab:kuaishou_dataset}
\end{table}

\textbf{DFI}:
Users usually only have impressions in a few domains, 
so their domain-related features in other domains are often missing, which makes it difficult to directly utilize the users' domain-related features for other domains.
The Automatic Domain Feature Integration uses domain features $\v_1, \v_2, ..., \v_D$ to enhance the prediction ability of each domain:
\begin{eqnarray} 
\z_d&=&\sum_{d=1}^{D}w_d h_1(\v_d),
\label{eqt:h1}
\end{eqnarray}
where $w_d$ is the personalized integration weight which is formulated as:
\begin{eqnarray} 
w_d=\frac{exp(\hat{w}_d)}{\sum_{d=1}^{D}exp(\hat{w}_d)},~~
\hat{w}_d=\frac{<h_2(\e_d),h_3(\v_d)>}{\sqrt{k}},
\label{eqt:h23}
\end{eqnarray}
where $<\cdot~, \cdot>$ represents the dot product.
$h_1(\cdot)$, $h_2(\cdot)$, and $h_3(\cdot)$ project the input information to one new unified vector space, which the dimension is $k$. In this paper, we use a simple MLP (Multi-Layer Perceptron) \cite{MLP} as $h_1(\cdot)$, $h_2(\cdot)$, and $h_3(\cdot)$ following previous work \cite{AITM}.
The key idea is similar to previous works \cite{nhfm2020neural,www2020modeling,difm2021modeling} and has been proven to be more effective.
The difference is that query $h_2(\e_d)$ is personalized for different samples, while key $h_3(\v_d)$ and value $h_1(\v_d)$ represent domain features, so they are consistent for all samples within the domain.

Finally, the prediction probability of each domain $d$ is:
\begin{eqnarray} 
\hat{y}_d=sigmoid(MLP(concat([\s_d, \u_d, \v_d, \z_d)])).
\end{eqnarray}

\subsection{Joint Optimization Strategy for MDL}
In the joint optimization phase, we need to cyclically feed mini-batch samples and minimize the \textit{cross-entropy} loss in each domain:
\begin{equation}
\L_d(\theta^s, \theta_d^u)=-\frac{1}{N}\sum^N_{(\x_d,y_d)\in\Domain_d}(\left(y_d\log\hat{y}_d+(1-y_d)\log(1-\hat{y}_d)\right),
\end{equation}
where $N$ is the number of samples, $y_d$ is the label of the $d$-th domain and $\theta^s$, $\theta_d^u$ is the shared and specific (unique) parameter sets in the MDL framework, respectively.
The $\theta^s$ contains the shared parameters in $E(\cdot)$, $f(\cdot)$, $h_1(\cdot), h_2(\cdot), h_3(\cdot)$ in Equations \eqref{eqt:e}, \eqref{eqt:f}, \eqref{eqt:h1}, \eqref{eqt:h23} cross all domains, and the other parameters is in $\theta_d^u$.
As shown in Algorithm \ref{alg:mdl}, the learning rate of the shared parameters $\theta^s$ is set to $\frac{1}{D}$ of the $\theta_d^u$ to balance the learning phase.

The framework is implemented using TensorFlow\footnote{https://www.tensorflow.org/} and trained through stochastic gradient descent over shuffled mini-batches with the Adam \cite{adam2015method} update rule.

\section{Experiments}
In this section, we evaluate our proposed framework by conducting extensive experiments on both industrial and public datasets. The experimental settings including dataset descriptions, reproducibility information, evaluation protocol, and baseline methods are introduced first. Then, we move on to our experimental results and analysis.
\begin{algorithm}[!t]
\caption{Joint Optimization Strategy for MDL}
\label{alg:mdl}
\begin{algorithmic}[1]
\REQUIRE training data $(\X, \Y)$, batch size $B$, learning rate $\gamma$.
\ENSURE The trained parameters $\theta^s$ and $\theta_d^u$.
\STATE Initializing $\theta^s$ and $\theta_d^u$ randomly.
\FOR{$epoch \leftarrow 1$ to $EPOCHS$}
     \FOR{$t \leftarrow 1$ to $STEPS$}
        \FOR{$d \leftarrow 1$ to $D$}
            \STATE Calculate gradient $\nabla J(\theta_d^u), \nabla J(\theta^s)$ based on $(\X_d, \Y_d)$
            \STATE $\theta_d^u = \theta_d^u - \gamma \nabla J(\theta_d^u)$
            \STATE $\theta^s = \theta^s - \frac{1}{D}\gamma \nabla J(\theta^s)$
        \ENDFOR
    \ENDFOR
\ENDFOR
\end{algorithmic}
\end{algorithm}

\subsection{Datasets}

Two datasets are used to validate our proposed framework. One is the real-world industrial dataset collected from \textit{Dianping}\footnote{https://www.dianping.com}, and the other is a publicly accessible dataset collected from \textit{Kuaishou}\footnote{https://www.kuaishou.com}. 
The tasks are all CTR prediction.
Tables \ref{tab:industry_dataset} and \ref{tab:kuaishou_dataset} show the statistical information of these two datasets. 
\begin{itemize}
    \item \textbf{Industrial dataset:} The industrial dataset is collected from the user log data of one of the most renowned local life information and trading platforms in China.
    We extract a total of 21 scenarios (domains) of different sizes in the application and denote them as \textit{\#A1} to \textit{\#A21} based on their CTR ranking. 
    The dataset comprises tens of millions of samples, encompassing user behavior, user attributes, and item attributes.
    We randomly split the dataset into training, validation, and test sets in a ratio of 8:1:1 according to the user id. 
    \item \textbf{Public dataset:} The public dataset is KuaiRand\footnote{https://github.com/chongminggao/KuaiRand} dataset \cite{kuairand}, which is collected from recommendation logs of the video-sharing mobile app, \textit{Kuaishou}. We use the KuaiRand-1K version and split domains according to the "tab" field. 
    In order to ensure the stability of evaluation results,
    we select the 6 largest domains (tab in [0,1,2,4,5,6]) and denote them as as \textit{\#B1} to \textit{\#B6}, respectively.
   According to the official guidelines, we designate the data from April 8 to 21, 2022, as the training set, the data from April 22 to May 8, 2022, as the test set, and the data from April 19 to 21, 2022, within the training set as the validation set.
\end{itemize}

\begin{table*}[!t]
  \centering
 \caption{The average AUC performance on the industrial dataset. Underlined results indicate the best baselines over each domain. ``*'' indicates that the improvement of the proposed \model~ is statistically significant compared with the best baselines at p-value $<$ 0.05 over paired samples t-test, and ``**'' indicates that the p-value $<$ 0.01.}
  \resizebox{\textwidth}{!}{
    \begin{tabular}{c|cccccccccccccccccccccc}
    \toprule
    \textbf{Model} & \textit{\textbf{\#A1}} & \textit{\textbf{\#A2}} & \textit{\textbf{\#A3}} & \textit{\textbf{\#A4}} & \textit{\textbf{\#A5}} & \textit{\textbf{\#A6}} & \textit{\textbf{\#A7}} & \textit{\textbf{\#A8}} & \textit{\textbf{\#A9}} & \textit{\textbf{\#A10}} & \textit{\textbf{\#A11}} & \textit{\textbf{\#A12}} & \textit{\textbf{\#A13}} & \textit{\textbf{\#A14}} & \textit{\textbf{\#A15}} & \textit{\textbf{\#A16}} & \textit{\textbf{\#A17}} & \textit{\textbf{\#A18}} & \textit{\textbf{\#A19}} & \textit{\textbf{\#A20}} & \textit{\textbf{\#A21}} & \textbf{avg} \\
    \midrule
    \textbf{MLP} & 0.6665  & 0.5808  & 0.6980  & 0.5557  & 0.6764  & 0.5699  & 0.5808  & 0.6363  & 0.6132  & 0.6658  & 0.6041  & 0.6331  & 0.5908  & 0.6476  & 0.5677  & 0.6274  & 0.6290  & 0.6715  & 0.6193  & 0.6326  & 0.5975  & 0.6221  \\
    \midrule
    \textbf{MMoE} & 0.7137  & 0.6574  & 0.7511  & 0.5487  & 0.7120  & 0.6286  & 0.6103  & 0.6792  & 0.6497  & 0.7058  & 0.6591  & 0.6433  & 0.6177  & 0.6588  & 0.6044  & 0.6419  & 0.6930  & 0.7193  & 0.6468  & 0.6635  & 0.6126  & 0.6579  \\
    \textbf{PLE} & 0.7192  & 0.6577  & \underline{0.7571}  & 0.5752  & 0.7118  & 0.6219  & 0.6105  & 0.6775  & 0.6554  & \underline{0.7118}  & 0.6519  & 0.6495  & 0.6123  & 0.6622  & 0.5941  & \textbf{\underline{0.6628}} & 0.6867  & 0.7170  & 0.6445  & 0.6592  & 0.6148  & 0.6597  \\
    \textbf{AITM} & 0.7036  & 0.6541  & 0.7491  & \textbf{\underline{0.5893}}  & 0.7296  & 0.6199  & 0.6189  & 0.6877  & 0.6365  & 0.7071 & 0.6591  & 0.6476  & 0.6242  & \textbf{\underline{0.6797}} & \underline{0.6169}  & 0.6462  & 0.6828  & 0.7136  & 0.6471  & 0.6704  & 0.6285  & 0.6625  \\
    \midrule
    \textbf{HMoE} & 0.7185  & 0.6525  & 0.7515  & 0.5704  & 0.7441  & 0.6266  & 0.6102  & 0.6746  & 0.6548  & 0.7079  & 0.6528  & 0.6040  & 0.6181  & 0.6697  & 0.5966  & 0.6559  & 0.6879  & 0.7106  & 0.6489  & 0.6631  & 0.6177  & 0.6589  \\
    \textbf{STAR} & 0.7158  & 0.6340  & 0.7501  & 0.5718  & 0.7579  & 0.6292  & 0.6099  & 0.6708  & 0.6536  & 0.7053  & 0.6543  & 0.6366  & 0.6161  & 0.6669  & 0.6007  & 0.6499  & 0.6775  & 0.7164  & 0.6469  & 0.6580  & 0.6192  & 0.6591  \\
    \textbf{PEPNet} & 0.7099  & 0.6585  & 0.7500  & 0.5644  & 0.7443  & 0.6209  & 0.6013  & 0.6807  & 0.6617  & 0.7063  & 0.6585  & 0.6364  & 0.6151  & 0.6628  & 0.6108  & 0.6439  & \underline{0.6942}  & 0.7188  & 0.6516  & 0.6623  & 0.6179  & 0.6605  \\
    \textbf{HiNet} & \textbf{\underline{0.7332}} & \underline{0.6587}  & 0.7409  & 0.5785  & \underline{0.7639}  & \underline{0.6363}  & \underline{0.6307}  & \underline{0.6980}  & \underline{0.6721}  & 0.6956  & \underline{0.6764}  & \textbf{\underline{0.6576}} & \underline{0.6293}  & 0.6787  & 0.6091  & 0.6496  & 0.6824  & \textbf{\underline{0.7213}} & \textbf{\underline{0.6561}} & \textbf{\underline{0.6740}} & \underline{0.6356}  & \underline{0.6704}  \\
    \textbf{DFEI} & 0.7316  & \textbf{0.6623**} & \textbf{0.7642*} & 0.5866 & \textbf{0.7820**} & \textbf{0.6434*} & \textbf{0.6383**} & \textbf{0.7050**} & \textbf{0.6729 } & \textbf{0.7137**} & \textbf{0.6831**} & 0.6449  & \textbf{0.6294 } & 0.6766  & \textbf{0.6329 } & 0.6558  & \textbf{0.6952 } & 0.7192  & 0.6535  & 0.6662  & \textbf{0.6423**} & \textbf{0.6761**} \\
    \bottomrule
    \end{tabular}%
    }
  \label{tab:result}%
\end{table*}%

\begin{table}[!t]
  \centering
  \caption{The summary of the hyper-parameters in all models.}
  \resizebox{\linewidth}{!}{
    \begin{tabular}{cc}
    \toprule
    Hyper-parameter & Value \\
    \midrule
    Optimizer & Adam \\
    Batch size & 512 \\
    Learning rate & 1e-3 \\
    L2 regularization & 1e-6 \\
    Embedding dimension & 16 \\
    Dimensions of layers in the MLP Tower and Expert& [128, 64, 32] \\ 
    Dropout rate in each layers of MLP & [0.1, 0.2, 0.3] \\
    Activation function in MLP & Relu \\
    \bottomrule
    \end{tabular}%
    }
  \label{tab:params}%
\end{table}%

\subsection{Reproducibility Information}
To ensure fair comparison and evaluate the generalization performance of different models, we adopt the same set of common hyper-parameters for all models on two datasets. These hyper-parameters are chosen based on empirical values and computational efficiency.
For both industrial and public datasets, the output dimension $k$ of the attention in  Equation \eqref{eqt:h23} is $32$, the decay coefficient $\alpha$ in Equation \eqref{eqt:avg} is $0.9$, we only fine-tune the hyper-parameters of $\alpha$ according to grid search on the validation set. 
Besides, the $h_1(\cdot)$, $h_2(\cdot)$ and $h_3(\cdot)$ are all single-layer MLP with dimension $32$ in Equations \eqref{eqt:h1}, \eqref{eqt:h23}. 
For a fair comparison, we try our best to ensure that the main architecture of different models is consistent. The summary of different models is shown in Table \ref{tab:params}.
We do not perform any dataset-specific tuning, except for the early-stopping on validation sets for all models, the experiments of all models are conducted with NVIDIA Tesla V100 GPU with 16G memory.

\begin{table}[!t]
  \centering
  \caption{The average AUC performance on the public dataset. Underlined results indicate the best baselines over each domain. ``*'' indicates that the improvement of the proposed \model~ is statistically significant compared with the best baselines at p-value $<$ 0.05 over paired samples t-test, and ``**'' indicates that the p-value $<$ 0.01.}
  \resizebox{\linewidth}{!}{
    \begin{tabular}{c|ccccccc}
    \toprule
    \textbf{Model} & \textit{\textbf{\#B1}} & \textit{\textbf{\#B2}} & \textit{\textbf{\#B3}} & \textit{\textbf{\#B4}} & \textit{\textbf{\#B5}} & \textit{\textbf{\#B6}} & \textbf{avg} \\
    \midrule
    \textbf{MLP} & 0.6505  & 0.7314  & 0.7658  & 0.7160  & 0.6770  & 0.7235  & 0.7107  \\
    \midrule
    \textbf{MMoE} & 0.6983  & 0.7325  & 0.7717  & 0.7203  & 0.6786  & 0.7284  & 0.7225  \\
    \textbf{PLE} & 0.6908  & 0.7347  & 0.7718  & 0.7223  & 0.6695  & 0.7207  & 0.7183  \\
    \textbf{AITM} & 0.6949  & \underline{0.7370}  & 0.7784  & 0.7220  & 0.6800  & \underline{0.7288}  & 0.7236  \\
    \midrule
    \textbf{HMoE} & 0.6995  & 0.7364  & 0.7801  & 0.7208  & 0.6857  & 0.7214  & 0.7240  \\
    \textbf{STAR} & 0.6989  & \underline{0.7370}  & 0.7771  & 0.7223  & 0.6831  & 0.7273  & 0.7243  \\
    \textbf{PEPNet} & 0.6948  & 0.7354  & 0.7837  & \textbf{\underline{0.7234}}  & 0.6853  & 0.7271  & 0.7250  \\
    \textbf{HiNet} & \underline{0.7051}  & 0.7364  & \underline{0.7842}  & 0.7224 & \textbf{\underline{0.6866}} & 0.7190  & \underline{0.7256}  \\
    \textbf{DFEI} & \textbf{0.7103**} & \textbf{0.7372} & \textbf{0.7931**} & 0.7195  & 0.6757  & \textbf{0.7328*} & \textbf{0.7281**} \\
    \bottomrule
    \end{tabular}%
    }
  \label{tab:addlabel}%
\end{table}%

\subsection{Evaluation Protocol}
Following existing works \cite{AITM, HiNet}, \textbf{AUC} (the area under ROC curve), a commonly used metric in ranking tasks, is adopted as the evaluation metric to compare the performance of the the baselines and the proposed \model~framework. 
To avoid interference caused by experimental fluctuations, all experiments were run five times with different random seeds. In the end, we presented the average AUC over the five runs and performed the paired samples t-test.
\begin{figure*}[!t]
\begin{center}
\includegraphics[width=1.0\linewidth]{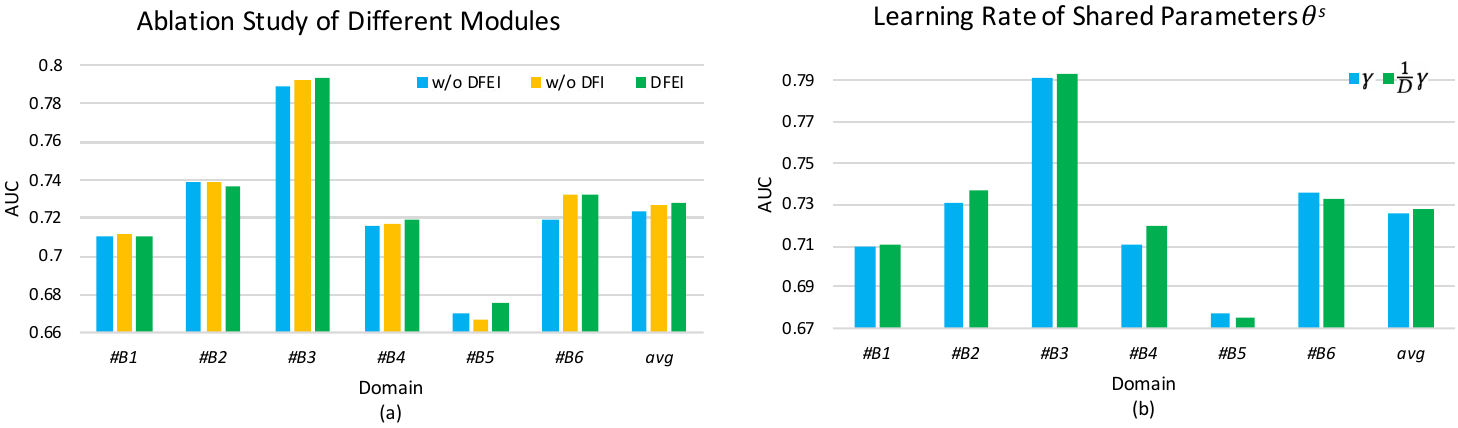}
\caption{Ablation Study of the \model~ modules and the learning rate of the shared parameters $\theta^s$ on the public dataset.}
\label{fig:ablation}%
\end{center}
\end{figure*}%
\subsection{Baseline Methods}
To demonstrate the effectiveness of our proposed framework, we compare it with several state-of-the-art baselines, including a single-task learning model, multi-task learning models, and multi-domain learning models. Below are more details about baselines:
\begin{enumerate}
    \item [\textbf{i)}] \textbf{Single-task learning model}
    \begin{itemize}
        \item [$\bullet$] \textbf{MLP}: This method utilizes a Multi-Layer Perceptron as the backbone and treats all domains as a single domain.
    \end{itemize}
    \item [\textbf{ii)}] \textbf{Multi-task learning models}
    \begin{itemize}
        \item [$\bullet$] \textbf{MMoE}\cite{MMoE}: This model adopts a gating mechanism to combine \textit{N} expert networks shared at the bottom to capture task-related information. Here, we consider different domains as distinct tasks and apply domain-specific towers and gating networks for each domain.
        \item [$\bullet$] \textbf{PLE}\cite{PLE}: This model based on MMoE, separates the experts into shared and task-specific to alleviate the negative transfer problem. 
        \item [$\bullet$] \textbf{AITM}\cite{AITM}: This model focuses on multi-task learning with sequential dependence, using a gating unit and an information transfer unit to adaptively learn what and how much information to transfer for different conversion stages of different audiences, thereby facilitating effective knowledge transfer.
    \end{itemize}
    \item [\textbf{iii)}] \textbf{Multi-domain learning models}
    \begin{itemize}
        \item [$\bullet$] \textbf{HMoE}\cite{HMoE}: This model is adapted from MMoE, utilizing a scenario-aware MOE structure to implicitly model distinctions and commonalities among different domains. 
        \item [$\bullet$] \textbf{STAR}\cite{STAR}: This model constructs a star topology incorporating both shared and domain-specific parameters, utilizing a single model to cater to all domains.
        \item [$\bullet$] \textbf{PEPNet}\cite{PEPNet}: This model introduces both parameter personalization and embedding personalization to accurately capture users' interest for multi-task recommendation in the multi-domain setting.
        \item [$\bullet$] \textbf{HiNet}\cite{HiNet}: This model applies a hierarchical information extraction structure with MOE module to model complex correlations between different domains and tasks.
    \end{itemize}
\end{enumerate}

\subsection{Performance Comparison}
The comparison of AUC scores among different methods on two datasets is reported in Tables \ref{tab:result} and \ref{tab:addlabel}, where the best score among all methods is marked in \textbf{boldface} and the best-performed baseline is \underline{underlined}. From the results, we have the following observations:

\begin{itemize}
    \item The simple MLP regards all domains as a single entity, failing to model the differences among domains. It employs entirely shared parameters for multi-domain learning, which makes the experimental performance fall short when compared to multi-task and multi-domain learning methods, particularly in the case of a large number of domains.
    \item Compared with MLP, multi-task learning models (MMoE, PLE, AITM) achieve better results on both datasets. We treat various domains as separate tasks, and this approach allows multi-task models to exhibit adaptability when dealing with multi-domain settings. 
    Besides, AITM focuses on multi-task learning with sequence dependence. When applied to multi-domain settings, it can establish a sequential relationship among different domains. Experimental results have indicated that domains ranked later in the sequence show a performance boost in the AITM model, compared to other multi-task models.
    Its overall performance is even better than some multi-domain baselines on the industrial dataset.
    However, multi-tasking methods focus more on modeling the relationships between tasks, thus ignoring domain differences when applied to multi-domain settings.
    \item Most multi-domain learning methods have modeled the similarities and differences among various domains, yielding positive results. 
    In addition, HiNet applies scenario-aware attentive network to effectively counteract the phenomenon of negative transfer, which makes it better capture and leverage the correlations among domains, leading to commendable performance.
    However, these models do not explicitly and automatically extract domain features for the large-scale multi-domain recommendation, and also do not explicitly address the issue of information utilization when users only have impressions in a limited number of domains, leading to sub-optimal performance.
    \item \model~ outperforms all baselines in average AUC score by a significant margin, showing superior predictive capabilities among domains and proving the effectiveness of combining automatic domain feature extraction and personalized integration modules. 
\end{itemize}


\begin{figure*}[ht]
 \begin{minipage}[t]{0.48\linewidth}
   \centering
   \includegraphics[width=\linewidth]{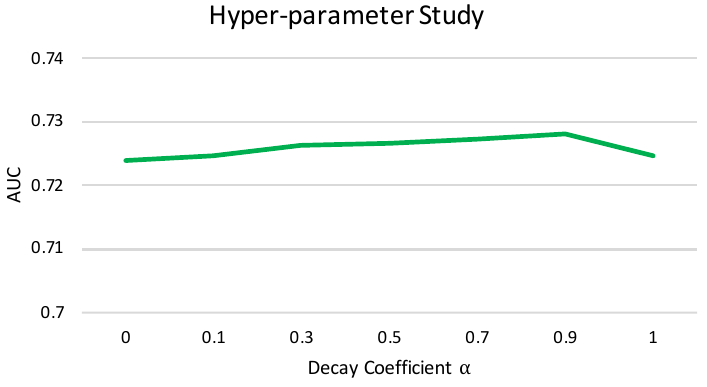}
   \caption{Hyper-parameter Study of the decay coefficient $\alpha$ on the public dataset. The metric is the average AUC of all domains.}
   \label{fig:params}
 \end{minipage}\hfill
 \begin{minipage}[t]{0.48\linewidth}
   \centering
   \includegraphics[width=\linewidth]{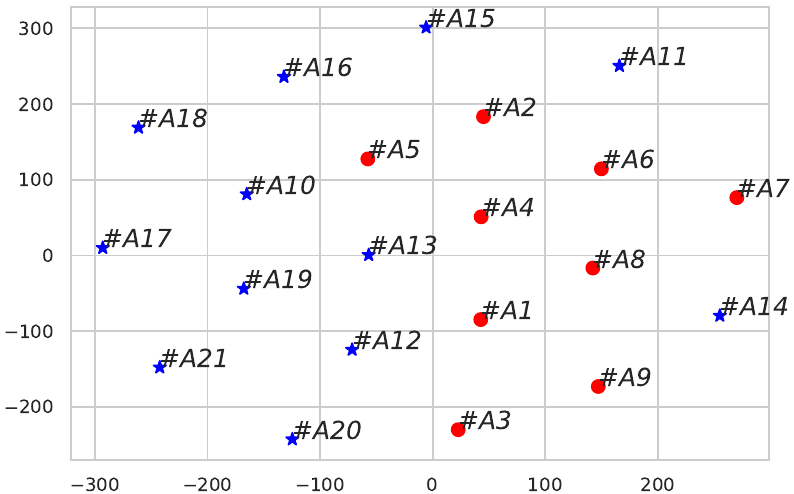}
   \caption{The t-SNE visualization of the final domain features $\v_1, ...,\v_D$ on the industrial dataset. The red dots represent CTR less than $5\%$, while the blue stars represent CTR greater than $5\%$ as shown in Table \ref{tab:industry_dataset}.}
   \label{fig:t-sne}
 \end{minipage}
\end{figure*}

\subsection{Ablation Study}
In this subsection, we perform the ablation study of the \model~ module and the learning rate of the shared parameters $\theta^s$ in Algorithm \ref{alg:mdl} on the public dataset.

Firstly, considering the impact of DFE and DFI in Figure \ref{fig:model}, we remove the DFEI (DFE + DFI, i.e., w/o DFEI) and  DFI (i.e., w/o DFI) respectively and the results are shown in Figure \ref{fig:ablation} (a).
Secondly, we study the impact of the learning rate of the shared parameters $\theta^s$, and the results are shown in Figure \ref{fig:ablation} (b).
From Figure \ref{fig:ablation}, we can obtain the following inspiring observations:
\begin{itemize}
    \item DFEI outperforms both w/o DFEI and w/o DFI in most cases, which suggests that the automatic domain feature extraction and personalized integration from different domains effectively enhance the prediction performance. Except for domains \textit{\#B2} and \textit{\#B5}, the AUC score of w/o DFI is usually higher than w/o DFEI. This suggests that employing automatic domain feature extraction is an effective strategy for illustrating the differences among domains, thereby enhancing the performance of \model. Some fluctuations occur in the results among domains, due to the significant disparity in data volumes. We consider the average AUC across all domains as our primary focus.
    \item With the learning rate scaling factor $\frac{1}{D}$ on the shared parameters $\theta^s$, the \model~ leads to higher average AUC scores.
    This is because in Algorithm \ref{alg:mdl}, each training step will be fed the samples of each domain, thus updating each domain-specific parameter $\theta^d$ once, while updating the shared parameter $\theta^s$ $D$ times. Therefore, using $\frac{1}{D}$ as the scaling factor will better balance the entire learning process.
    
\end{itemize}
From the above results, we could see that the learning rate scaling factor of the shared parameters and the \model~ module can indeed improve the performance of multi-domain recommendation.

\subsection{Hyper-parameter Study}

In order to study the impact of the hype-parameter decay coefficient $\alpha$ in Equation \eqref{eqt:avg}, we perform the hyper-parameter study on the public dataset. 
As shown in Figure \ref{fig:params}, we vary the decay coefficient as [0, 0.1, 0.3, 0.5, 0.7, 0.9, 1] and report the average AUC performance on the test set. 
When $\alpha$ is small, it will forget more historical information and record more of the latest batch information, resulting in poor performance. As $\alpha$ gradually increases, more historical information is preserved and accumulated. Finally, the behavior of each individual user is transformed into an aggregation of all user behaviors within the domain, and the best performance is achieved when $\alpha=0.9$.

\subsection{Visualization Study}
In order to determine if the final domain features ($\v_1, ...,\v_D$ in Equation \eqref{eqt:avg}) have successfully captured specific information across different domains, we utilize t-SNE (t-distributed Stochastic Neighbor Embedding \cite{t-sne}) to visualize them in Figure \ref{fig:t-sne}.
Due to the limited domain number of the public dataset and the lack of statistical significance, our visualization study is conducted on the industrial dataset.
From the visualization, we have the following interesting findings: Domains with a CTR less than $5\%$ ($\#A1 - \#A9$) are concentrated in the area with x coordinates greater than 0,
while domains with a CTR greater than $5\%$ ($\#A10 - \#A21$) are mostly concentrated in the area with x coordinates less than 0.
This indicates that the domain features have learned the CTR information of the domain, and the CTR is directly correlated with the target label. 
Furthermore, the domain features are higher-level feature representations, and the CTR information is merely a dimension indicating differences between domains.

\section{Conclusion}
In this paper, we proposed an Automatic Domain Feature Extraction and Personalized Integration (\model) framework for the large-scale multi-domain recommendation. The proposed framework could automatically extract and personalize the integration of domain features for each user, which can yield more accurate conversion identification. 
Experimental results on large-scale multi-domain datasets demonstrate significant improvement compared with state-of-the-art baseline models.

Several directions are available for future research in the area. Firstly, the \model~ could be applied as a plug-and-play method to other advanced models.
Secondly, combining the MTL and MDL may bring additional gains.



\bibliographystyle{ACM-Reference-Format}
\balance
\bibliography{sample-base}

\end{document}